# Mid-Epidemic Forecasts of COVID-19 Cases and Deaths: A Bivariate Model applied to the UK

Peter Congdon, School of Geography, Queen Mary University of London, Mile End Rd, London E1 4NS, UK; p.congdon@qmul.ac.uk

**Abstract** *Background*. The evolution of the COVID-19 epidemic has been accompanied by efforts to provide comparable international data on new cases and deaths. There is also accumulating evidence on the epidemiological parameters underlying COVID-19. Hence there is potential for epidemic models providing mid-term forecasts of the epidemic trajectory using such information. The effectiveness of lockdown or lockdown relaxation can also be assessed by modelling later epidemic stages, possibly using a multiphase epidemic model. *Methods*. Commonly applied methods to analyze epidemic trajectories or make forecasts include phenomenological growth models (e.g. the Richards family of densities), and variants of the susceptible-infected-recovered (SIR) compartment model. Here we focus on a practical forecasting approach, applied to interim UK COVID data, using a bivariate Reynolds model (for cases and deaths), with implementation based on Bayesian inference. We show the utility of informative priors in developing and estimating the model, and compare error densities (Poisson-gamma, Poisson-lognormal, Poisson-logStudent) for overdispersed data on new cases and deaths. We use cross-validation to assess medium term forecasts. We also consider the longer term post-lockdown epidemic profile to assess epidemic containment, using a two phase model. *Results*. Fit to interim mid-epidemic data shows better fit to training data and better cross validation performance for a Poisson-logStudent model. Estimation of longer term epidemic data after lockdown relaxation, characterised by protracted slow downturn and then upturn in cases, casts doubt on effective containment. *Conclusions*. Many applications of phenomenological models have been to complete epidemics. However, evaluation of such models based simply on their fit to observed data may give only a partial picture, and cross-validation against actual trends is also valuable. Similarly, it may be preferable to model incidence rather than cumulative data, though this raises questions about suitable error densities for modelling often erratic fluctuations. Hence there may be utility in evaluating alternative error assumptions.

## 1. Introduction

Epidemic forecasts have been an essential element in policy decisions regarding the COVID-19 epidemic, such as lockdown imposition and relaxation. Forecasting has been assisted by well-organized efforts to provide international data on new cases and deaths. These include the daily updated comparative data provided by the European Centre for Disease Prevention and Control (ECDC) (https://www.ecdc.europa.eu/en/publications-data), and monitoring profiles provided by John Hopkins University [1]. There is also a growing literature providing evidence on the parameters of the COVID-19 infection (for example, case fatality ratios, serial intervals, etc). Hence the potential occurs for epidemic models that are applicable to routinely collected data, that make use of accumulated evidence, and can provide forecasts for epidemics observed at mid-stage. Policy decisions in many countries (imposition of lockdowns, and later relaxation) have been made based on trends in observed numbers of cases and deaths, while admitting these may be subject to measurement error - for example, identified cases may understate total numbers infected; there are fluctuations in daily new cases due to variation in daily testing; and there may be COVID-19 diagnostic errors.

Commonly applied approaches to quantitative modelling of aggregate epidemic data differ in the data inputs they require, their assumptions, their estimability, and in their scope for practical application to making forecasts. Commonly applied methods include phenomenological growth models [2], such as the Richards family of densities [3], and variants of the susceptible-infected-recovered (SIR) compartment model [4]. Phenomenological models are parameterised in terms of epidemic trajectories and provide estimates of crucial epidemiological parameters [5], while avoiding the complexity of more formal models of disease transmission which can be difficult to estimate and provide projections [4,6]. As mentioned by Wu et al [7], phenomenological models are "particularly suitable when significant uncertainty clouds the epidemiology of an infectious disease". By contrast, as noted in [8], compartmental transmission models may be based on untested assumptions such as random mixing between all individuals in a given population, may be sensitive to starting assumptions, and may provide estimates that differ considerably between models. Such models often rely on preset parameters, which may mean prediction uncertainty is understated. They may also be complex to specify when an epidemic has more than one phase.

Here we focus on a practical forecasting approach using routinely available data on new cases and deaths (from ECDC) to estimate parameters of a bivariate version of the Reynolds model. Implementation is based on Bayesian inference principles, incorporating accumulated evidence on relevant parameters via informative priors. The operation and utility of this approach is demonstrated using data on new cases and deaths in the United Kingdom (UK), with a focus on predictive accuracy for 20-day ahead forecasts of cases and deaths, based on mid-epidemic data. Other policy relevant parameters such as the effective reproduction ratio are also estimated in analysis of longer term epidemic data.

## 2. Related Research

A number of studies, with widely differing methodologies, have been made which forecast different aspects of the COVID epidemic or related health care need. In its impact on policy making in the UK and US, perhaps most influential has been the Imperial College model [9]. This is based on microsimulation with transmission through contacts between susceptible and infectious individuals in various settings, or randomly in the community, depending on spatial distance between contacts. A number of epidemic parameters (e.g. incubation periods and basic reproduction numbers) are preset. Forecasts are provided for deaths and hospital beds. Also providing forecasts across countries is the model of the Institute for Health Metrics and Evaluation [8]. This has no underlying representation of epidemic dynamics, but is based on fitting a hierarchical parametric model for observed cumulative death rates in different countries, and then projecting these forward.

Various types of time series forecast have also been made, using ARIMA models [10, 11], exponential smoothing [12], or autoregression in cases; for example, the study by Johndron et al [13] postulates daily deaths as a lagged function of earlier new cases. Autoregression in cases, with potential for short term forecasting, is illustrated (for foot and mouth disease) by the first order autoregressive model of Lawson et al [14], while (for multiple spatial units) the model of Bracher and Held [15] specifies a first order autoregression based on the mean incidence in adjacent areas.

Regarding phenomenological models, the study by Zhao et al [5] exemplifies their application to epidemic forecasting, though to the Zika epidemic in 2015 rather than the COVID-19 epidemic. Applications of such models to COVID-19 include [16] and [17].

## 3. **Methods**

### 3.1 Phenomenological Models

Basic phenomenological models for epidemic trajectories include the logistic, Gompertz, and Rosenzweig, which have been the basis for a range of generalizations [2, 3]. Application of the logistic model to COVID-19 is exemplified by the studies of Batista [18] and Shen [19]. For time $t$, the logistic model for new cases $C'(t)$ and cumulative cases $C(t)$ is

$$C'(t) = rC(t)\left[1 - \frac{C(t)}{K}\right], \qquad (1)$$

$$C(t) = \frac{K}{\left[1+e^{-r(t-\tau_L)}\right]},$$

where $K$ is the maximum number of cases (final epidemic size), $r > 0$ measures the intensity of exponential growth in cases in the early epidemic phase, and $\tau_L$ is the inflection point where new cases are highest. The Richards model [20] modifies the logistic incidence function to

$$C'(t) = rC(t)\left[1 - \left(\frac{C(t)}{K}\right)^a\right], \qquad (2)$$

with solution

$$C(t) = \frac{K}{\left[1+e^{-r(t-\tau)}\right]^{1/a}}.$$

The parameter $a > 0$ modifies the incidence decline phase of the logistic, that is measures the extent of deviation from the standard logistic curve. The turning point $\tau$, when incidence peaks, is obtained when $C(t)$ equals $K(1 + a)^{-1/a}$ [21]. The peak incidence is important for the healthcare planning, for example aligning the forecast peak with hospital bed capacity [22].

Other commonly used models are the Gompertz model [23] with

$$C'(t) = rC(t)\log[K/C(t)],$$

while the Rosenzweig [24] has

$$C'(t) = rC(t)\Big[\Big(\frac{C(t)}{K}\Big)^a - 1\Big].$$

The incidence function represented by $C'(t)$ can be used to define mean incidence in statistical likelihoods for new cases data. Thus time series of incidence counts can often be satisfactorily modelled as a Poisson, with means defined by $C'(t)$ functions [25, 26]. Similarly the cumulative cases function $C(t)$ can be used to define mean epidemic size in models for cumulative case counts [24, 27].

It may be preferable to estimate epidemic model parameters by analyzing new cases rather than cumulative cases, as the latter have serially correlated measurement error, leading to understatement of parameter uncertainty [28]. As stated in [28], "independence of sequential measurement errors, ... is clearly violated when observations are accumulated through time". However, estimation using new cases and deaths puts a much greater focus on how to deal with stochastic variation in the data. For daily data, fluctuations in new events may be considerable, whereas cumulative cases and deaths are usually smooth functions.

While for smaller epidemics, a Poisson density for mean incidence may be applied, for larger epidemics such as COVID-19, a negative binomial density is often preferred, both because of large incidence counts, and to represent often erratic incidence fluctuations that lead to overdispersion relative to the Poisson [28]. Other overdispersed versions of the Poisson can be achieved by mixing the Poisson with a suitable density (e.g. a lognormal density) [29,30], and this may be beneficial in detecting unusual observations.

3.2 Model Specification: Poisson Overdispersion

Consider the Richards model parameters. Let $c_t$ and $C_t$ denote incidence and cumulative incidence counts at times $t = 1,\ldots,T$ (days in the case of COVID data from ECDC). We condition on the first case or cases (i.e. the first observation), and take incidence at time $t$ as a function of cumulative cases at $t-1$, so that for a Poisson model we have

$$c_t \sim Poisson(\mu_t)$$

$$\mu_t = rC_{t-1}[1-(C_{t-1}/K)^a], \qquad t = 2,\ldots,T.$$

In practice many epidemic datasets are overdispersed relative to the Poisson, and many studies adopt a negative binomial model instead. We can specify an overdispersed model (including the negative binomial) by introducing multiplicative random effects [29, eqn 2], such that the Poisson means for incidence are specified by

$$\mu_t = rC_{t-1}[1-(C_{t-1}/K)^a]\epsilon_t, \qquad (3)$$

where the $\epsilon_t$ are positive random effects. For the Poisson-gamma model (equivalent to a Negative Binomial), the $\epsilon_t$ are gamma distributed with mean 1, namely

$$\epsilon_t \sim Gamma(\lambda, \lambda),$$

where $1/\lambda$ is the overdispersion parameter mentioned by [28]. Note that the assumed parameterisation of the gamma density with random variable $x$ is

$$p(x|a,b) = \frac{b^a}{\Gamma(a)} x^{a-1} e^{-bx}.$$

Other options in (3) are to take $u_t = \log(\epsilon_t)$ as normally distributed [29]

$$u_t \sim Normal(0, \sigma_\epsilon^2),$$

or Student t distributed [31],

$$u_t \sim Student - t(0, \sigma_\epsilon^2, \nu)$$

where $\nu$ is a degrees of freedom parameter. These two options define the Poisson-Lognormal (PLN) and Poisson-log-Student (PLS) options respectively [30]. The PLN and PLS representations may provide a more robust alternative to the Poisson-gamma [31, 32, 33, 34, 35], as their tails are heavier than for the gamma distribution, and have been found to be better at accommodating outliers (such as daily "spikes" in an epidemic application).

### 3.3 A Joint Model for New Cases and New Deaths

In the analysis below we apply a bivariate estimation with both new cases and deaths modelled using the Richards specification. Thus denote $d_t$ and $D_t$ as new and cumulative deaths at time t. The joint likelihood for an overdispersed Poisson model for both outcomes then specifies

$$c_t \sim Poisson(\mu_{ct}),$$

$$d_t \sim Poisson(\mu_{dt}),$$

$$\mu_{ct} = r_c C_{t-1}[1 - (C_{t-1}/K_c)^{a_c}]\epsilon_{ct}, \quad (4)$$

$$\mu_{dt} = r_d D_{t-1}[1 - (D_{t-1}/K_d)^{a_d}]\epsilon_{dt}, \qquad t = 2, \ldots, T. \quad (5)$$

For a Bayesian application, we need to specify prior densities, or priors for short, for the parameters. For the epidemic size parameter $K_c$, a diffuse prior confined to positive values, such as a diffuse gamma density - for example $Ga(1, \varepsilon)$ or $G(\varepsilon, \varepsilon)$ , with $\varepsilon$ small - was found to lead to convergence problems. As noted in [18],"... in the early stage, the logistic curve follows an exponential growth curve, so the estimation of $K$ is practically impossible". This difficulty persists when an epidemic is past its peak but early in a downturn. However, Batista [18] mentions a relationship (for the logistic model) between successive cumulative case counts that may assist in providing a prior for $K_c$ . Specifically for three points spaced $m$ time units apart, one may obtain the relationship for a point estimator of $K_c$ , namely

$$K_c^e = \frac{C_{t-m}(C_{t-2m}C_{t-m} - 2C_{t-2m}C_t + C_{t-m}C_t)}{(C_{t-m}^2 - C_t C_{t-2m})}.$$

One may use this point estimator to define a prior mean for $K_c$ in the Richards model

(which is a generalisation of the logistic). Specifically, one may take a lognormal density prior for $K_c$ , with $log(K_c^e)$ as mean, and a suitable variance, such that the prior is still relatively diffuse. For example, suppose $K_c^e$ is 250000, and the variance in the lognormal is set at 1. Then the 97.5 percentile for the lognormal prior is 1.73 million.

In the bivariate specification (for new cases and deaths jointly), we seek to share prior information between outcomes. One option for the prior on $K_d$ (the final death total) is as a function of $K_c$ , namely

$$K_d = \Phi K_c,$$

where $\Phi$ is a form of case fatality ratio (CFR). An informative prior for $\Phi$ could be based on the COVID experience in similar countries, or on experience of epidemics of similar diseases. Considering the first option, and an appropriate prior for analysing UK data, an informative prior for $\Phi$ could be provided by the case fatality ratio across the European Union (the UK being no longer an EU member). International information on case fatality is provided at https://ourworldindata.org/mortality-risk-covid#the-current-case-fatality-rate-of-covid-19.

Alternatively one may link $K_c$ and $K_d$ using both the point estimator $K_c^e$ and a case fatality ratio, namely $K_d^e = \Phi K_c^e.$ Then a lognormal density prior can be taken for $K_d$ , with $log(K_d^e) = log(K_c^e) + \log(\Phi)$ as the mean, and a suitable variance such that the prior is still relatively diffuse.

Another possible prior to link ultimate cases and deaths would involve a time series in time-specific case fatality ratios, such as an autoregression $\phi_t \sim N(\rho\phi_{t-1}, \sigma_\phi^2)$ , with $\phi_t$ estimated from cumulative data on deaths and cases. Some analyses of epidemics show that the CFR early in an epidemic may underestimate later values [36], in which case the prior on $\Phi$ may be constrained to exceed the final $\phi_t$ based on observed data. With regard to COVID-19 not all epidemics fit this pattern, with the US (for example) showing a decline in CFRs at later epidemic stages. There is also evidence that the infection to mortality ratio (a more precise measure than the CFR) has fallen [37].To allow for such a scenario, the prior for $\Phi$ could be centred at the last observed $\phi_t,$ rather than constrained to exceed it.

Joint priors on other parameters could be considered, for example, bivariate normal priors on the logs of $r_c$ and $r_d$ , or on the logs of $a_c$ and $a_d$ . In the empirical analysis below we focus on priors linking the final epidemic and death total parameters, $K_c$ and $K_d$ , as these are an important influence on forecasts.

### 3.4 Medium Term Forecasts

Many applications of phenomenological models are to historic data on epidemics, where the epidemic has run to its full extent. Here we consider applications to incomplete epidemics (e.g. epidemics observed to their mid point or early in the downturn), and to forecasts using such data. Forecasts at an intermediate point within the observation span are of interest in themselves for policy purposes. However, they can also be used in comparative model evaluation by using cross-validation, with only some data used for estimating the model, and some held out for validation.

Thus suppose the training sample is formed by observations up to time $M < T$, while the $F$ subsequent observations at times $t = M+1,\ldots,M+F$ (where $M+F \leq T$) are used as a validation sample. Predictions $c_{new,M+1}$ and $d_{new,M+1}$ for new cases and deaths at time $M+1$ are based on observed cumulative counts $C_M$ and $D_M$ . As usual in Bayesian inference, predictions are obtained as replicate data sampled from the posterior predictive densities $p(c_{new}|Y) = \int p(c_{new}|Y,\theta)d\theta$ and $p(d_{new}|Y) = \int p(d_{new}|Y,\theta)d\theta$ where $Y = (c,d)$ are data on new cases and deaths, and $\theta$ are parameters in the joint model of section 3.3 [38].

Predicted cumulative counts at time $M+1$ are then obtained as $C_{new,M+1} = C_{new,M} + c_{new,M+1}$ and $D_{newM+1} = D_{new,M} + d_{new,M+1}.$ Predicted new cases and deaths at $M+2$ , $c_{new,M+2}$ and $d_{newM+2},$ are then sampled from the appropriate phenomenological model form, with predicted new cases and deaths at $M+2$ based on $C_{new,M+1}$ and $D_{new,M+1}.$ Cumulated cases and deaths at $M+2$ are then obtained by adding predicted new cases and deaths for $M+2$ to $C_{new,M+1}$ and $D_{new,M+1}.$ This process is continued until time $M+F.$

Fit can be assessed by whether credible intervals for predictions in the cross-validation period include actual incidence and new deaths. Also relevant are probabilities of over or under-prediction. For example, consider predicted new cases $c_{new,s,t}$ for the validation period $t = M+1,\ldots,M+F,$ and for MCMC samples $s = 1,\ldots,S$ , and let average new predicted cases during the validation period (the average over $F$ days) at iteration $s$ be denoted $\bar{c}_{new,s,T+1:T+F}$ . We want to compare average predicted new cases with average actual new cases, $\bar{c}_{T+1:T+F}$ , during the validation period. Probabilities of overprediction can be obtained from binary indicators

$$O_s^{(c)} = I(\bar{c}_{new,s,T+1:T+F} > \bar{c}_{T+1:T+F}), \qquad s = 1,\ldots,S,$$

where $I(A) = 1$ if condition $A$ is true, and $I(A) = 0$ otherwise. Thus at each iteration we compare average new cases (modelled) during the validation period with actual average new cases.

Probabilities of overprediction for new cases, $\omega_c$, are estimated as $\sum_{s=1}^{S} O_s^{(c)}/S.$ Probabilities of underprediction can be obtained as $1-\omega_c.$ A satisfactory prediction would have $0.05<\omega_c<0.95,$ with $\omega_c$ over 0.95 indicating a high probability of overprediction, while $\omega_c$ under 0.05 indicates a high probability of underprediction. Underprediction means under-forecasting of future cases and may lead to incorrect inferences regarding epidemic control, as it implies a lessening in incidence earlier than actually occurred.

### 3.5 Later Epidemic Stages and Effective Reproduction Ratios

Strategic decisions regarding containment of the COVID-19 epidemic, in the UK and other countries, have depended on trends in new infections and deaths, but also on the effective reproduction rate. Thus in the UK the choice on whether or not to relax the initial COVID lockdown restructions was based on five criteria, with two being numeric: first, "a sustained and consistent fall in daily death rates", and second that the "rate of infection is decreasing to manageable levels", meaning that the effective reproduction ratio is demonstrably below 1. The reproduction rate may also become especially relevant at later epidemic stages (post-lockdown), after a downturn from the initial peak and after lockdown measures have been relaxed. Here the concern is to prevent a resurgence of infection, indicated by an upturn in $R_t$ .

In the case of a protracted downturn, with new cases remaining non-neglible, the concern is especially that there may be a resurgence in cases, and possibly also deaths, at some point. This is colloquially known as a "second wave", and in some European countries (e.g. France, Spain) there have been pronounced second waves in the COVID-19 epidemic. Such a resurgence indicates use of a multiphase model [39], with a second phenomenological model applied to data after a switch-point between epidemic regimes.

Consider a model for new cases only. The two phenomenological models (before and after the switch-point) have distinct parameters. The switch-point is taken as a latent parameter, $\kappa,$ such that for a two-phase Richards model

$$c_t \sim Poisson(\mu_t)$$

$$\mu_t = I(t<\kappa)r_1C_{t-1}[1-(C_{t-1}/K_1)^{a_1}] \quad + \quad I(t \geq \kappa)r_2C_{t-1}[1-(C_{t-1}/K_2)^{a_2}]$$

where $I(A)=1$ when condition $A$ is true, and $0$ otherwise. The parameter $\kappa$ can be assigned a uniform prior (on a positive interval) or a positive valued prior, such as an exponential density. If there is a second wave upturn in deaths also, then a bivariate model can be used, with switch points later for deaths than cases, due to delays in mortality upturns following incidence upturns. A three phase model would have two switch-points, with mean

$$\mu_t = I(t<\kappa_1)r_1C_{t-1}[1-(C_{t-1}/K_1)^{a_1}] \quad + \quad I(\kappa_2>t \geq \kappa_1)r_2C_{t-1}[1-(C_{t-1}/K_2)^{a_2}] + I(t \geq \kappa_2)r_3C_{t-1}[1-(C_{t-1}/K_3)^{a_3}].$$

An estimator of the effective reproduction rate $R_t$ at time $t$ is based on predictions from the phenomenological model, possibly multiphase, and from an estimate of the serial interval density. The serial interval is the time between symptom onset in an infected subject and symptom onset in the infectee. The serial interval density can discretized in the form of weights $\rho_j$, applied to serial interval lengths (in days) up to a maximum $J$. These can be used to estimate effective reproduction ratios $R_t$ within a phenomenological model to analyse new cases; see the papers [25], [40] and [41]. Thus $R_t$ can be estimated as

$$R_t = c_{new,t} / \sum_{j=0}^{J} \rho_j c_{new,t-j},$$

where $c_{new,t}$ are predicted new case data from the phenomenological model.

## 4. Model Application

We consider the application of the above methods to UK data on new cases and deaths from 1st February 2020 (when the first two cases of COVID-19 in the UK were reported according to ECDC). Observations are assigned dates as in the ECDC data. Figures 1a and 1b show daily trends in these outcomes up to 8th August 2020, with erratic fluctuations apparent in both outcomes. However, there is a broad downward trend in both outcomes from days 70 to 80, though with a more protracted decline as opposed to the initial upturn. Figures 2a and 2b show the relatively smooth evolution of cumulative cases and deaths.

### 4.1 Medium Term Forecasts

For medium term forecasts, we focus on the Richards bivariate model (section 3.3) and compare three alternative error assumptions as discussed in section 3.2: Poisson-Gamma (PG), Poisson-lognormal (PLN) and Poisson-log-Student (PLS). The Student density is represented by a scale mixture of normals [42]. Cross-validation estimations are made at three points $M < T$. Thus we consider twenty day ahead forecasts at three different stages of the UK COVID-19 epidemic. For the first cross validation, estimations are based on training data up to day $M = 80$ with $F = 20$ (i.e. the cross validation period consists of days $81$ to $100$). Cross validation estimations with $F = 20$ are also made for $M = 100$ and for $M = 120$, with forecast accuracy based on comparing forecasts with hold out data for days $M + 1, \ldots, M + F$.

### 4.2 Model Assumptions

For prior densities on the unknowns in the medium term forecasts, exponential priors with mean 1 are assumed on $r_c$, $r_d$, $a_c$ and $a_d$. For the precision $1/\sigma_\epsilon^2$ in the PLN and PLS options, and for the parameter $\lambda$ in the Poisson-gamma model, a gamma prior $Gamma(1, 0.001)$ is assumed. For the PLS option, we take $\nu = 4$ as a preset option. The degrees of freedom can be difficult to estimate for relatively small datasets, and the option of the preset value $\nu = 4$ is a robust option [43, 44]. For the maximum cases (epidemic size) parameter $K_c$ we assume as lognormal prior, centred at $log(K_c^e),$ as discussed in section 3.2.

For the maximum deaths parameter $K_d,$ we assume $K_d = \Phi K_c$ where the prior for $\Phi$ is a beta with mean defined by the EU-wide case fatality $\phi_{EU}$ with total prior count $C$ set at 5, $C = 5$ . Thus the prior on $\Phi$ is $Beta(C\phi_{EU}, C(1 - \phi_{EU})).$ For example, on 20-04-2020 (at day t=80), the EU wide case fatality rate was 0.101, and the prior is set at $\Phi \sim Beta(0.505, 4.495).$

4.3 Model Fit and Estimation
Fit to the observed training data is assessed using the Watanabe--Akaike information criterion (WAIC) [45]. Cross-validation fit is assessed by the probabilities of overprediction $\omega_c$ and $\omega_d$ , and by predictive coverage: whether the credible intervals for predicted cases and deaths (averages over each validation period) include the actual averages. Estimation uses the BUGS program [46], with posterior estimates based on the second halves of two chain runs of 100,000 iterations, and convergence assessed using Brooks-Gelman-Rubin criteria [47].

4.2 Modelling Later Epidemic Stages
To provide a longer term perspective on epidemic containment, we apply the best performing model to the full set of observations as at 8th August 2020 (T=190) when the first epidemic peak had passed. At this point there were 309 thousand cumulative cases and 46500 cumulative deaths. New cases and deaths (as 7 day moving averages) had reached maxima of 4850 and 950 respectively when the UK epidemic peaked in April. As a result of lockdown measures imposed in late March, by July daily new cases averaged just over 500 daily. However, lockdown relaxations from July were accompanied by the risk of resurgence. In that regard, an upturn was apparent with new cases in late July and early August averaging over 800 daily (see Figure 3), though deaths continued to fall, averaging around 50 per day in early August.

To model the full time series, and since there is no upturn in deaths, we focus on new cases only. A two phase model is applied with shift parameter $\kappa$ assigned an exponential density with mean 150. Priors on the parameters of the second phase Richards model are as before for the exponential ascent and logistic modifier parameters. For $K_2$ we assume $K_2 = \eta K_1$ with $\eta$ assigned an exponential prior with mean 1. Of policy interest here is epidemic containment, as summarised by the effective reproduction ratio: specifically is this ratio consistently below 1, and its 95% credible interval also entirely below 1.

To provide estimates of the effective reproduction ratio, we use accumulated evidence on COVID serial intervals from five studies [48,49,50,51,52]. A gamma density on the serial interval (SI) is assumed, and information on mean and standard deviation of the SI, or on quantiles of SI, is converted to gamma density parameters; for use of SI quantiles in this regard, see [53]. Large samples (of a million) from each the five densities are taken, and parameters of the pooled gamma density are estimated from the pooled sample of five million; the pooled density has shape 1.38 and rate 0.36. This density is then converted to a discretized form (with 16 bins) to provide an informative prior on the SI to the model of White and Pagano [54], which updates the prior serial interval density using new case data for the UK. The paper [55] recommends that the initial, approximately exponential, epidemic phase be used in estimation, and we use UK new case data up to time 24-04-2020, when UK new cases peaked [see 55, page 3]. The updated mean serial interval is estimated as 3.5 days with standard deviation 3.1. The discretized serial interval is estimated as in [54], with $J = 16$.

## 5. Results and Discussion

5.1 Table 1 compares parameter estimates from the observed (training) data under the three error assumptions for the three cross-validation analyses at M=80, M=100 and M=120. It should be noted that predictions of cases and deaths should be based not on the posterior mean or median parameter values in Table 1, but on sampled replicate data at each iteration. These are based on sampling new data from the Poisson means (4) and (5), and from the Richards model parameter profile at each particular iteration. The predicted values of cases and deaths are very close to actual values: for example, for the PLS model at M=120, the average absolute deviation (over $t = 2, \ldots, 120$) between actual new cases and predicted new cases is under 1. Table 2 compares the WAIC fits to the observed data under the nine options, while criteria regarding 20 day ahead forecasts are shown in Table 3.

5.2 Mid Term Forecasts.

Table 1 shows broad consistency between the three distributional options in terms of estimated final epidemic size $K_c$ and eventual death total $K_d$. The posterior density of these parameters may be skewed, with posterior mean exceeding median. For the PLN and PLS options the estimate of $K_c$ increases as $M$ does. This reflects the protracted nature of the UK downturn in cases. Posterior mean estimates of the turning points $\tau_c$ and $\tau_d$ vary slightly and tend to be higher for $M = 100$ and $M = 120$, but for both outcomes and all $M$ values are between 72 and 86. Turning point estimates for new cases $\tau_c$ are also mostly higher under the PLS option.

Table 2 shows that the PLS option has better fit, with lower WAIC values. Hence its estimates of epidemic size and turning points are preferred, and provide a better description of the slow decline in cases from their peak. Table 3 shows generally better cross-validation performance for the PLS option. As discussed above, satisfactory prediction would have $0.05 < \omega_c < 0.95,$ and $0.05 < \omega_d < 0.95,$ with $\omega_c$ or $\omega_d$ over 0.95 indicating overprediction, and with $\omega_c$ or $\omega_d$ under 0.05 indicating underprediction. Both PG and PLN options show underprediction for higher values of $M$ , and in policy terms, may be misleading in suggesting a faster decline in cases and deaths than actually occurred. By contrast for $M = 100$ and $M = 120$ , the 95% interval for predicted average cases under the PLS model comfortably includes the actual average new cases.

5.3 Longer Term Scenario

Estimates of policy relevant parameters also come from the longer term scenario when the log-Student Richards model is applied to UK COVID new case data up to day T=190 (8th August, 2020). The shift point in the two phase model is estimated as 179.1 with 95% interval from 179.0 to 179.3.

Figure 4 plots out the posterior mean estimated effective reproductive ratios (5 day moving averages), distinguishing those estimates significantly above 1. The estimates hover around 1 throughout June and the first half of July, but in late July/early August tend to exceed 1. The impression from this is that success in fully effective containment is in doubt. The estimates of the reproduction ratio, and their path through June and July, are similar to those for the UK available at https://epiforecasts.io/covid/posts/national/united-kingdom/#national-summary and based on the methods in [56].

**6. Conclusion**

Many applications of phenomenological models have been to complete epidemics. However, evaluation of such models based simply on their fit to observed data may give only a partial picture. Also relevant to epidemic model assessment, particularly for policy application, is the accuracy of medium term forecasts for incomplete epidemics. Arguably, evaluation in this case is better done using a cross-validation approach, where only some of the observed data are used to estimate parameters and a hold out sample can assess the accuracy of forecasts. The analysis here has shown that fit to training data and the cross validation fit are consistent in their choice of preferred model option.

Many applications of phenomenological models have also been to cumulative cases data. However, to quote [57] on cholera epidemics, "The [incidence] data from Zimbabwe show multiple peaks and other features characteristic of heterogeneously mixed populations [...]. The practice of fitting epidemic models to cumulative incidence curves rather than incidence curves can obscure these features while also violating statistical assumptions of independence between fitted data points". A focus on incidence rather than cumulative incidence does therefore bring into sharper focus the question of adequately representing often erratic fluctuations in the observed series, apparent in UK COVID data on new cases and deaths.

For larger counts, and overdispersed count data, the literature has mainly focussed on the negative binomial as the solution to such variability. So far as the authors are aware, the literature does not contain any assessment of alternative distributional choices to the Poisson-gamma mixture (negative binomial). Hence the contribution of this paper rests on a comparison of the negative binomial against alternative Poisson mixture models [30]. The analysis suggests these may usefully be considered as alternatives to the negative binomial, and may both improve fit to actual observations and provide more accurate forecast performance.

There is also value in applying phenomenological models to later stages of incomplete epidemics, especially (in the case of COVID-19) after lockdown relaxation where there may be second waves of infection. The above analysis has therefore applied the Poisson-log Student to late epidemic data for the UK,with a focus on the effective reproduction rate or effective reproduction ratio. As mentioned in [58], "the COVID-19 pandemic has shown that the effective reproduction rate of the virus $R_t$ is a crucial determinant not only of public health, but also of public policy". There are a number of ways of estimating this quantity, including novel approaches such as using Google mobility data [59]. Here an analysis using estimates of $R_t$ based on a two-phase Richards model suggest a upturn in transmission in the UK.

Competing Interests: The authors declare no competing interests.

**Table 1 Estimations according to Timing of Cross-Validation Period (M is days after start of epidemic)**

**M=80**

| | Poisson-gamma | | | | Poisson-lognormal | | | | Poisson-log Student | | | |
|---|---|---|---|---|---|---|---|---|---|---|---|---|
| Parameter | Mean | 2.5% | Median | 97.5% | Mean | 2.5% | Median | 97.5% | Mean | 2.5% | Median | 97.5% |
| $K_c$ | 373800 | 151800 | 210500 | 1172000 | 170200 | 135900 | 153200 | 289000 | 198800 | 135100 | 177200 | 336600 |
| $K_d$ | 32920 | 25070 | 32080 | 45560 | 33220 | 24180 | 32390 | 46770 | 35780 | 17950 | 29060 | 91460 |
| $r_c$ | 0.31 | 0.22 | 0.27 | 0.58 | 0.24 | 0.19 | 0.23 | 0.30 | 0.12 | 0.08 | 0.13 | 0.18 |
| $r_d$ | 0.82 | 0.44 | 0.75 | 1.51 | 0.80 | 0.38 | 0.67 | 1.81 | 0.29 | 0.17 | 0.28 | 0.46 |
| $a_c$ | 0.31 | 0.07 | 0.32 | 0.60 | 0.54 | 0.21 | 0.55 | 0.89 | 1.06 | 0.53 | 1.06 | 1.79 |
| $a_d$ | 0.13 | 0.05 | 0.12 | 0.26 | 0.14 | 0.04 | 0.13 | 0.33 | 0.75 | 0.20 | 0.63 | 1.75 |
| $\phi$ | 0.13 | 0.03 | 0.15 | 0.25 | 0.21 | 0.10 | 0.20 | 0.31 | 0.18 | 0.09 | 0.16 | 0.40 |
| $\tau_c$ | 76.2 | 72 | 75 | 80 | 72.5 | 70.0 | 72.0 | 80.0 | 76.1 | 73.0 | 75.0 | 80.0 |
| $\tau_d$ | 74.5 | 71 | 74 | 79 | 74.7 | 71.0 | 75.0 | 80.0 | 75.8 | 71.0 | 76.0 | 80.0 |

**M=100**

| | Poisson-gamma | | | | Poisson-lognormal | | | | Poisson-log Student | | | |
|---|---|---|---|---|---|---|---|---|---|---|---|---|
| Parameter | Mean | 2.5% | Median | 97.5% | Mean | 2.5% | Median | 97.5% | Mean | 2.5% | Median | 97.5% |
| $K_c$ | 271040 | 243490 | 268820 | 310190 | 262700 | 245700 | 261100 | 284000 | 295800 | 223800 | 276400 | 453200 |
| $K_d$ | 37395 | 35097 | 37274 | 40538 | 37040 | 34470 | 36920 | 40340 | 38740 | 32290 | 37610 | 53060 |
| $r_c$ | 0.30 | 0.22 | 0.30 | 0.38 | 0.26 | 0.22 | 0.26 | 0.33 | 0.18 | 0.13 | 0.17 | 0.27 |
| $r_d$ | 1.51 | 0.63 | 1.21 | 4.01 | 1.27 | 0.48 | 1.06 | 2.88 | 0.55 | 0.22 | 0.49 | 1.12 |
| $a_c$ | 0.25 | 0.17 | 0.24 | 0.42 | 0.28 | 0.20 | 0.28 | 0.37 | 0.46 | 0.20 | 0.46 | 0.83 |
| $a_d$ | 0.07 | 0.02 | 0.06 | 0.13 | 0.08 | 0.02 | 0.07 | 0.17 | 0.20 | 0.06 | 0.15 | 0.62 |
| $\phi$ | 0.14 | 0.12 | 0.14 | 0.16 | 0.14 | 0.13 | 0.14 | 0.16 | 0.14 | 0.08 | 0.13 | 0.20 |
| $\tau_c$ | 79.4 | 78 | 79 | 82 | 79.0 | 78 | 79 | 80 | 83.1 | 78 | 82 | 97 |
| $\tau_d$ | 76.2 | 75 | 76 | 77 | 76.1 | 75 | 76 | 77 | 77.6 | 75 | 77 | 83 |

**M=120**

| | Poisson-gamma | | | | Poisson-lognormal | | | | Poisson-log Student | | | |
|---|---|---|---|---|---|---|---|---|---|---|---|---|
| Parameter | Mean | 2.5% | Median | 97.5% | Mean | 2.5% | Median | 97.5% | Mean | 2.5% | Median | 97.5% |
| $K_c$ | 302800 | 295900 | 302300 | 312100 | 303100 | 293900 | 302700 | 317100 | 349100 | 282000 | 321300 | 549100 |
| $K_d$ | 40670 | 39670 | 40630 | 41890 | 40370 | 39410 | 40330 | 41560 | 41050 | 38580 | 40770 | 45020 |
| $r_c$ | 0.38 | 0.30 | 0.36 | 0.51 | 0.34 | 0.25 | 0.32 | 0.49 | 0.40 | 0.11 | 0.45 | 0.80 |
| $r_d$ | 1.87 | 0.65 | 1.50 | 5.38 | 1.57 | 0.67 | 1.52 | 2.82 | 0.94 | 0.34 | 0.78 | 2.35 |
| $a_c$ | 0.16 | 0.11 | 0.16 | 0.21 | 0.18 | 0.11 | 0.18 | 0.24 | 0.27 | 0.04 | 0.30 | 0.69 |
| $a_d$ | 0.05 | 0.01 | 0.04 | 0.11 | 0.05 | 0.02 | 0.04 | 0.10 | 0.09 | 0.03 | 0.08 | 0.22 |
| $\phi$ | 0.13 | 0.13 | 0.13 | 0.14 | 0.13 | 0.13 | 0.13 | 0.14 | 0.12 | 0.07 | 0.13 | 0.15 |
| $\tau_c$ | 81.0 | 80 | 81 | 82 | 81.2 | 81 | 81 | 82 | 85.6 | 82 | 84 | 99 |
| $\tau_d$ | 77.3 | 77 | 77 | 78 | 77.1 | 77 | 77 | 78 | 77.9 | 77 | 78 | 79 |

**Table 2 Fit to Training Data (WAIC Criterion)**

| | Poisson-gamma | Poisson-lognormal | Poisson-log Student |
|---|---|---|---|
| | | M=80 | |
| Cases | 605.0 | 600.0 | 564.1 |
| Deaths | 365.7 | 367.4 | 350.4 |
| Total | 970.7 | 967.4 | 914.5 |
| | | M=100 | |
| Cases | 851.6 | 845.8 | 808.8 |
| Deaths | 562.0 | 560.1 | 550.1 |
| Total | 1413.6 | 1405.9 | 1358.9 |
| | | M=120 | |
| Cases | 1097.0 | 1086.3 | 1042.3 |
| Deaths | 743.3 | 741.4 | 731.5 |
| Total | 1840.4 | 1827.7 | 1773.8 |

## Table 3 Criteria for Out-sample Predictions (M is Number of Days in Training Sample; F=20 in all cases)

### M=80

| | Poisson-gamma (PG) | | | | Poisson-lognormal (PLN) | | | | Poisson-log Student (PLS) | | | |
|---|---|---|---|---|---|---|---|---|---|---|---|---|
| Criterion | Mean | 2.5% | Median | 97.5% | Mean | 2.5% | Median | 97.5% | Mean | 2.5% | Median | 97.5% |
| Average Daily Cases | 4357 | 1818 | 3686 | 7407 | 2214 | 1078 | 1881 | 4875 | 3118 | 1043 | 2869 | 5652 |
| Average Daily Deaths | 609 | 344 | 603 | 904 | 613 | 303.8 | 613.4 | 921.8 | 587 | 0 | 514 | 1471 |
| $\omega_c$ | 0.41 | | | | 0.03 | | | | 0.16 | | | |
| $\omega_d$ | 0.34 | | | | 0.38 | | | | 0.40 | | | |
| Actual Daily Averages in period (M+1,M+F) | | | | | | | | | | | | |
| New Cases | 4857 | | | | | | | | | | | |
| New Deaths | 662 | | | | | | | | | | | |

### M=100

| | Poisson-gamma (PG) | | | | Poisson-lognormal (PLN) | | | | Poisson-log Student (PLS) | | | |
|---|---|---|---|---|---|---|---|---|---|---|---|---|
| Criterion | Mean | 2.5% | Median | 97.5% | Mean | 2.5% | Median | 97.5% | Mean | 2.5% | Median | 97.5% |
| Average Daily Cases | 2346 | 1523 | 2353 | 3173 | 1975 | 1384 | 1947 | 2658 | 2630 | 622 | 2565 | 4340 |
| Average Daily Deaths | 265 | 184 | 265 | 351 | 224.8 | 135.8 | 221.5 | 332.8 | 275 | 52 | 276 | 510 |
| $\omega_c$ | 0.09 | | | | 0.00 | | | | 0.31 | | | |
| $\omega_d$ | 0.06 | | | | 0.03 | | | | 0.32 | | | |
| Actual Averages in period (M+1,M+F) | | | | | | | | | | | | |
| Cases | 3039 | | | | | | | | | | | |
| Deaths | 330 | | | | | | | | | | | |

### M=120

| | Poisson-gamma (PG) | | | | Poisson-lognormal (PLN) | | | | Poisson-log Student (PLS) | | | |
|---|---|---|---|---|---|---|---|---|---|---|---|---|
| Criterion | Mean | 2.5% | Median | 97.5% | Mean | 2.5% | Median | 97.5% | Mean | 2.5% | Median | 97.5% |
| Average Daily Cases | 1132 | 1006 | 1129 | 1274 | 1133 | 963.2 | 1136 | 1320 | 1310 | 604 | 1358 | 1835 |
| Average Daily Deaths | 129 | 89 | 129 | 170 | 117.9 | 77.6 | 117.6 | 160.4 | 135 | 37 | 134 | 236 |
| $\omega_c$ | 0.00 | | | | 0.00 | | | | 0.42 | | | |
| $\omega_d$ | 0.00 | | | | 0.00 | | | | 0.07 | | | |
| Actual Averages in period (M+1,M+F) | | | | | | | | | | | | |
| Cases | 1506 | | | | | | | | | | | |
| Deaths | 216 | | | | | | | | | | | |

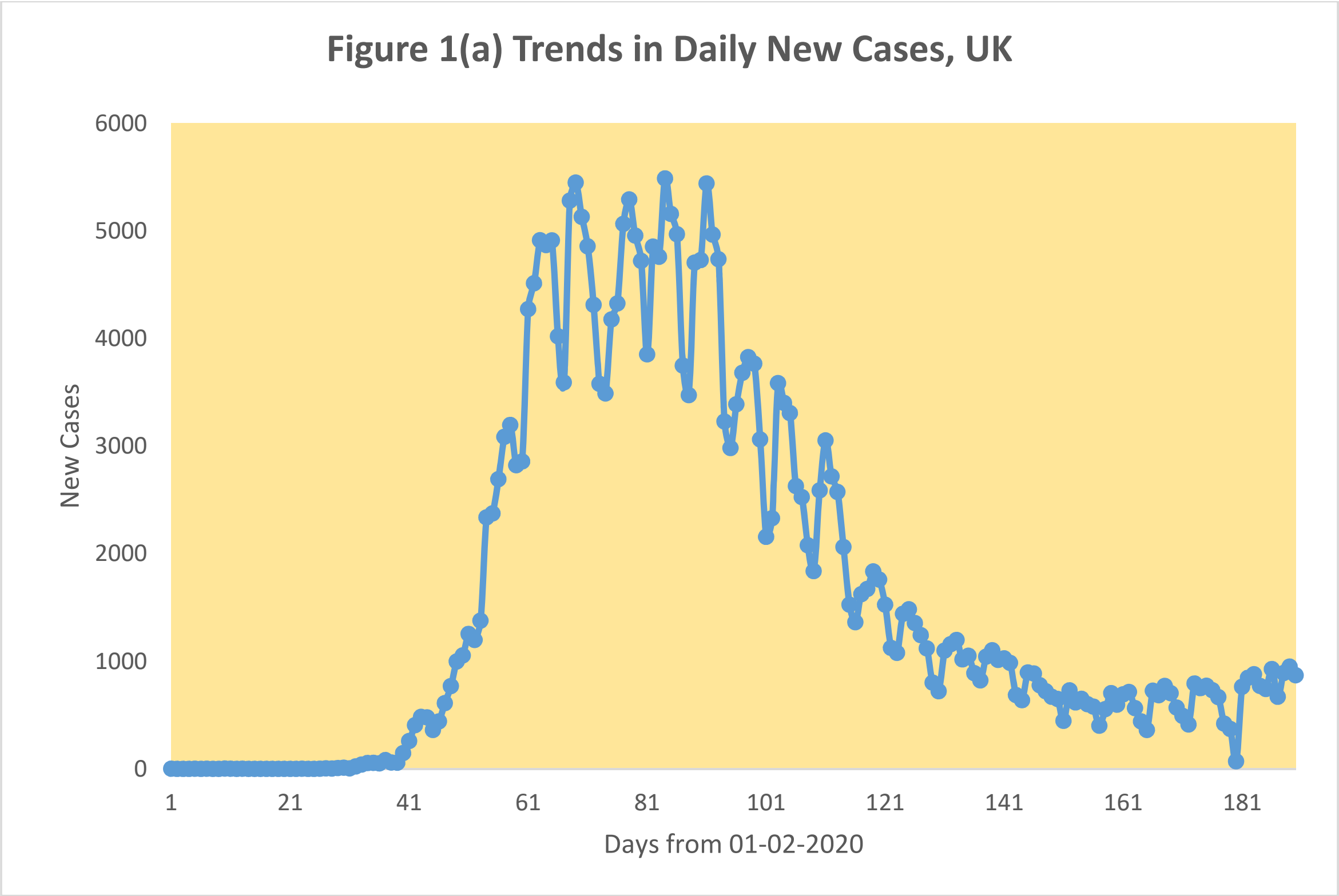
Figure 1(a) Trends in Daily New Cases, UK
New Cases
6000
5000
4000
3000
2000
1000
0
1
21
41
61
81
101
121
141
161
181
Days from 01-02-2020

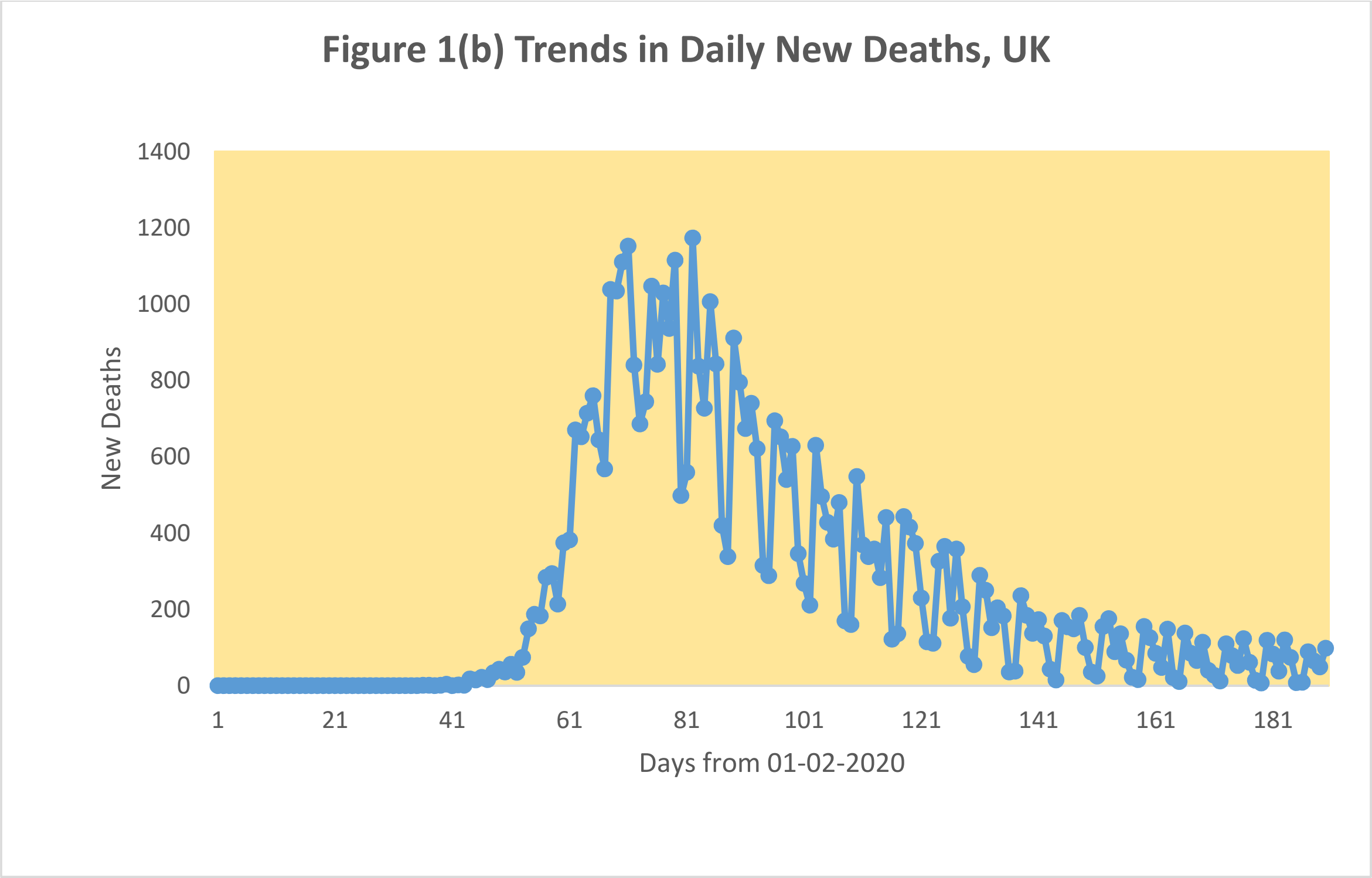
Figure 1(b) Trends in Daily New Deaths, UK
New Deaths
1400
1200
1000
800
600
400
200
0
1
21
41
61
81
101
121
141
161
181
Days from 01-02-2020

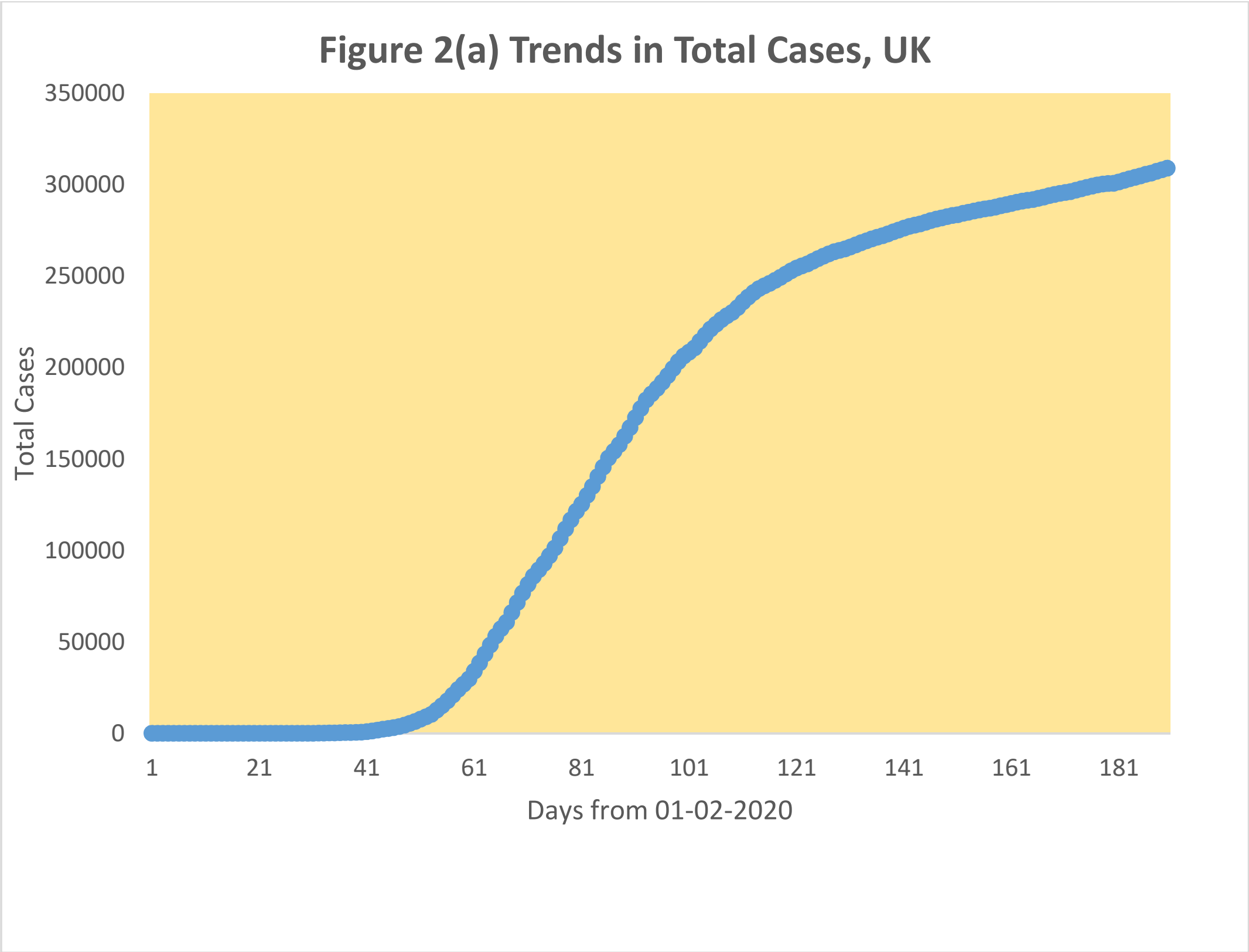

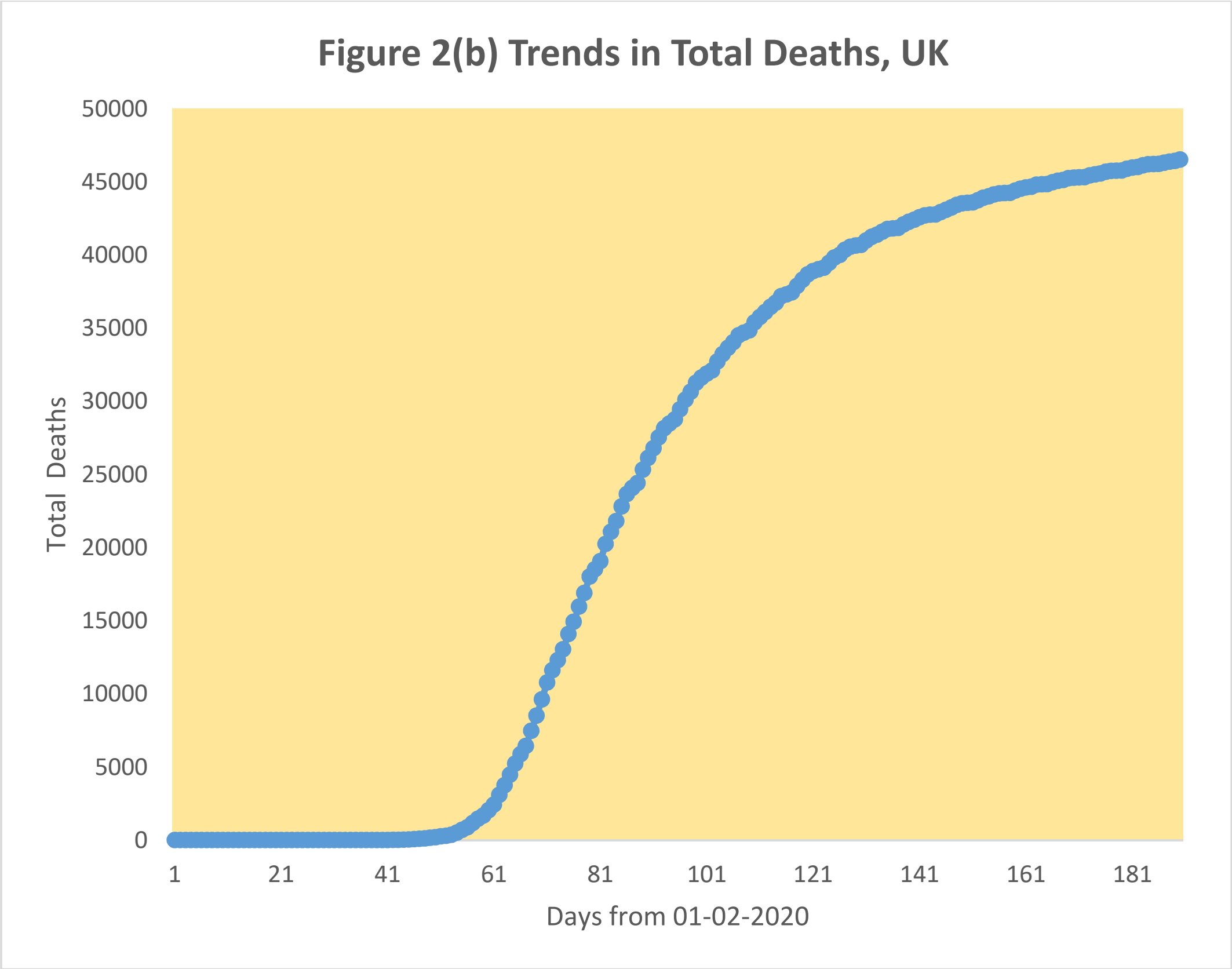

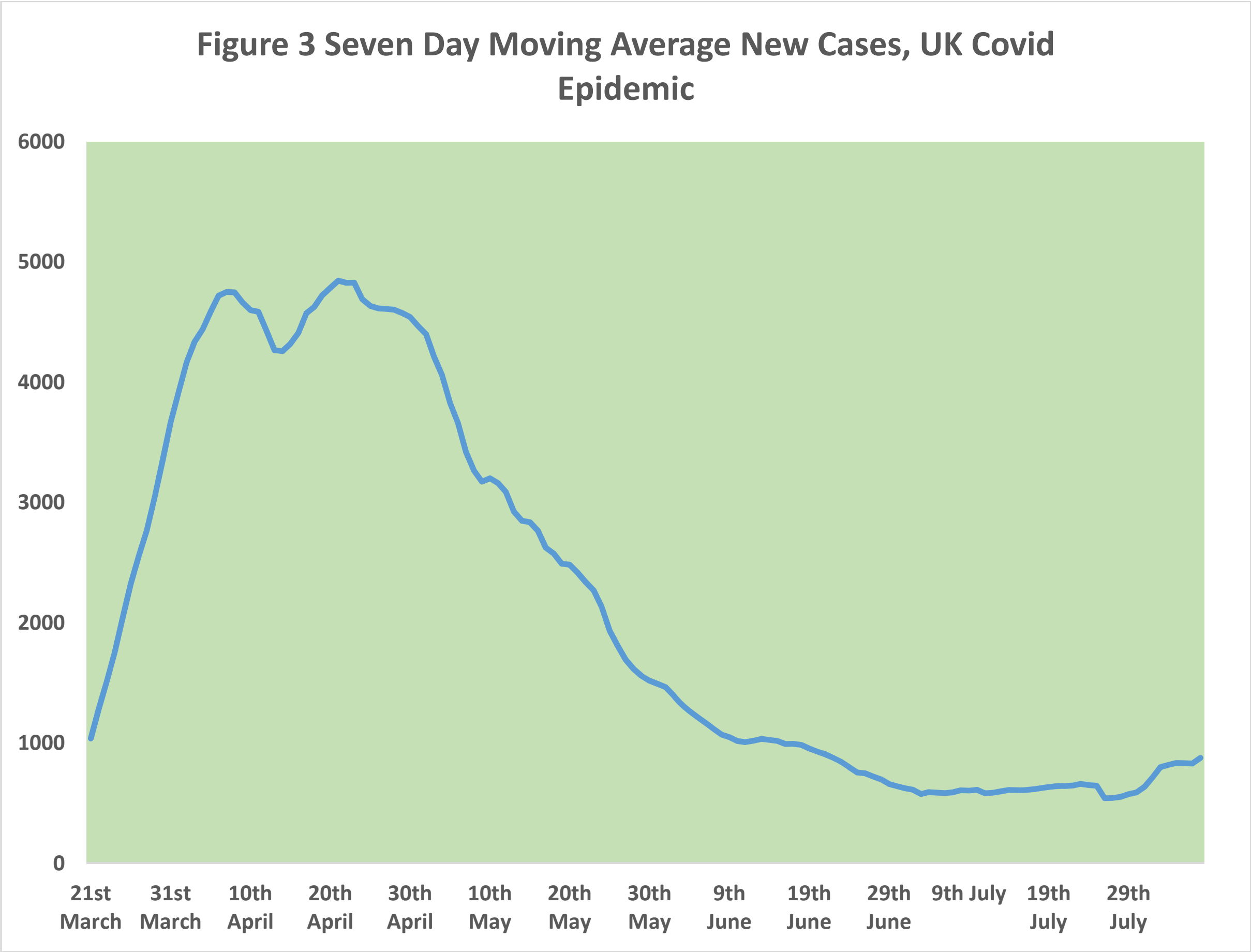
Figure 3 Seven Day Moving Average New Cases, UK Covid Epidemic
6000
5000
4000
3000
2000
1000
0
21st March
31st March
10th April
20th April
30th April
10th May
20th May
30th May
9th June
19th June
29th June
9th July
19th July
29th July

Figure 4 Reproduction Ratios and Significance

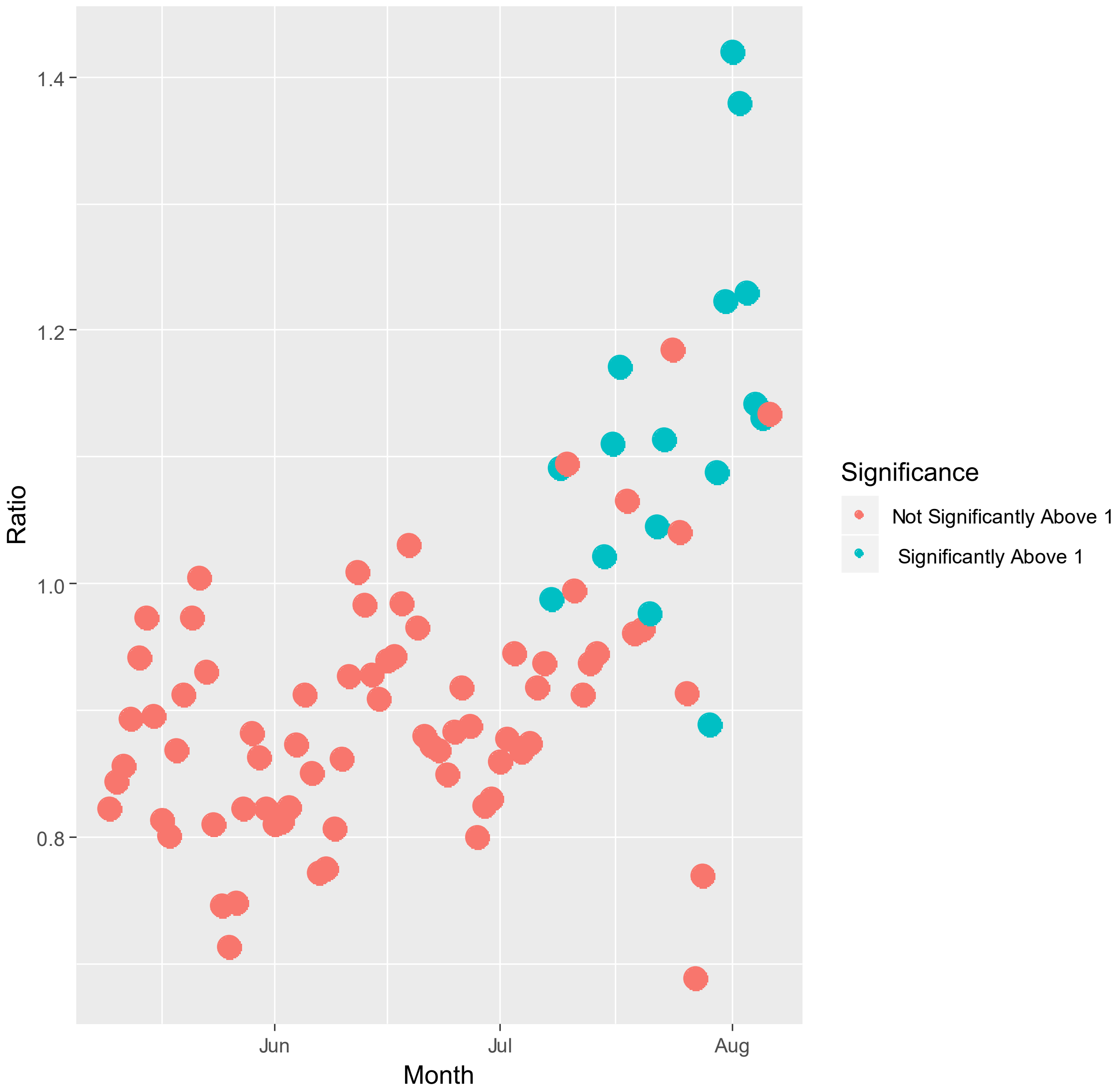